\documentclass[preprint,12pt]{elsarticle}

\usepackage{amssymb}
\usepackage{braket,amsmath,tabularray}
\usepackage{multirow}

\journal{Neurocomputing}

\begin{document}

\begin{frontmatter}



\title{Plugin Speech Enhancement: A Universal Speech Enhancement Framework Inspired by Dynamic Neural Network}


\author[1]{Yanan Chen\corref{fn1}}
\ead{chenyanan@chinamobile.com}

\author[1]{Zihao Cui\corref{fn1}}
\ead{cuizihaoyjy@chinamobile.com}

\author[1]{Yingying Gao}
\ead{gaoyingying@chinamobile.com}

\author[1]{Junlan Feng}
\ead{fengjunlan@chinamobile.com}

\author[1]{Chao Deng}
\ead{dengchao@chinamobile.com}

\author[1]{Shilei Zhang\corref{cor1}}
\ead{zhangshilei@chinamobile.com}

\cortext[fn1]{These authors contributed equally to this work and should be considered co-first authors.}
\cortext[cor1]{Corresponding author}

\affiliation[1]{organization={China Mobile Research Institute},
            city={Beijing},
            country={China}}
\begin{abstract}
The expectation to deploy a universal neural network for speech enhancement, with the aim of improving noise robustness across diverse speech processing tasks, faces challenges due to the existing lack of awareness within static speech enhancement frameworks regarding the expected speech in downstream modules. These limitations impede the effectiveness of static speech enhancement approaches in achieving optimal performance for a range of speech processing tasks, thereby challenging the notion of universal applicability. The fundamental issue in achieving universal speech enhancement lies in effectively informing the speech enhancement module about the features of downstream modules.
In this study, we present a novel weighting prediction approach, which explicitly learns the task relationships from downstream training information to address the core challenge of universal speech enhancement. We found the role of deciding whether to employ data augmentation techniques as crucial downstream training information. This decision significantly impacts the expected speech and the performance of the speech enhancement module.
Moreover, we introduce a novel speech enhancement network, the Plugin Speech Enhancement (Plugin-SE). The Plugin-SE is a dynamic neural network that includes the speech enhancement module, gate module, and weight prediction module. Experimental results demonstrate that the proposed Plugin-SE approach is competitive or superior to other joint training methods across various downstream tasks.
\end{abstract}


\end{frontmatter}


\section{Introduction}
Single-channel speech enhancement (SE) stands as a foundational technique crafted to enhance speech quality and intelligibility, which is to suppress noise while keep the speech undistorted as much as possible \cite{Loizou2013speech}.
In the realm of speech processing, various approaches have been utilized to improving noise robustness. These approaches include data augmentation and the incorporation of independent enhancement models, such as those used in telecommunication \cite{telecom_2023moore,wang2022hgcn,cui2023semi}, automatic speech recognition (ASR) \cite{NR_ASR_2013Seltzer,NR_ASR_2022Hu,chang2022end}, speaker verification (SV) \cite{NR_E_SR_2019Zhou,Kim2023PASPA,Shon2019VoiceIDLS} and emotion recognition \cite{NR_ER_2019Trian}.
Furthermore, the SE method can be applied to expand the speech corpus for a large speech language model \cite{latif2023sparks,gslm,speechlm,yang2023uniaudio} by preprocessing noisy speech.

\begin{figure}[tb]
    \centering
    \centerline{\includegraphics[width=10cm]{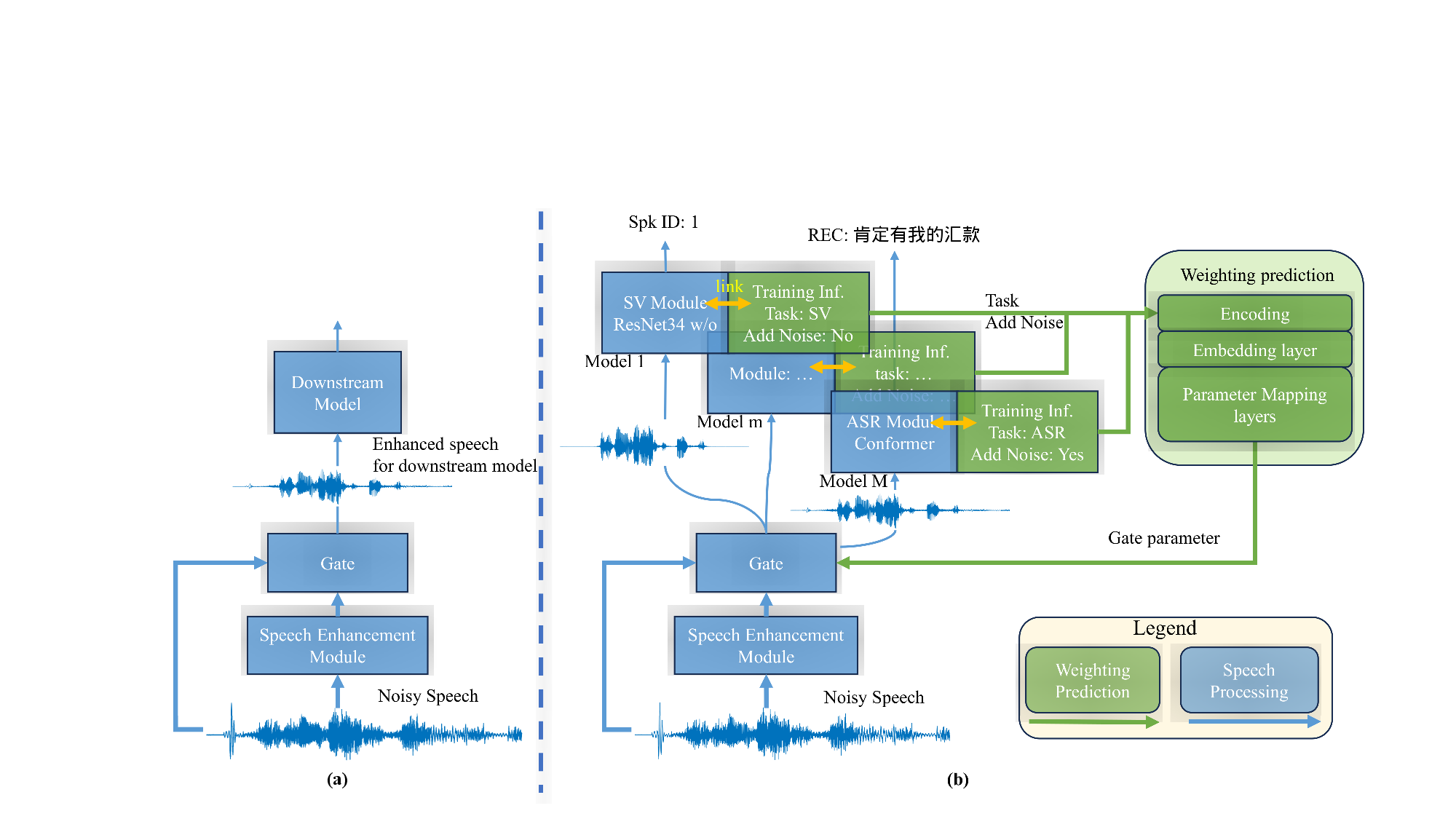}}
\caption{The inference stage of static speech enhancement and proposed plugin speech enhancement. (a) the static speech enhancement, (b) the plugin speech enhancement, as a kind of dynamic speech enhancement.}
\label{fig:Plugin-SE concept}
\end{figure}

For the sake of achieving high performance, as depicted in figure \ref{fig:Plugin-SE concept} (a), these SE networks devote efforts to designing training targets for specific environmental-robust tasks. However, this specialization can lead to a lack of flexibility in model learning across a range of environmental-robust tasks, necessitating the deployment of an equal number of dedicated downstream models. This proves impractical, particularly when computational resources are limited.

The primary approach to address this challenge is through multi-task learning \cite{trinh2022unsupervised, du2020self, ma2018modeling} or dynamic neural network \cite{han2022dynamic}. Despite some research efforts in multi-task speech enhancement, achieving performance comparable to specialized speech enhancement networks involving distinct downstream models remains a challenge. This challenge arises from the significant differences in the expected enhanced speech for various downstream models.
In this paper, we categorize these downstream models into \textbf{noise-sensitive models} and \textbf{environmental-robust models}. The environmental-robust model refers to downstream model training that utilizes data augmentation techniques, especially with noise injection, which is only sensitive to artificial noise, while noise-sensitive models do not use the noise injection technique, which is sensitive to all kinds of noise. For noise-sensitive downstream models, the primary of speech enhancement \cite{telecom_2023moore, wang2022hgcn, JesperK2023PSD} is to minimize speech distortion. In contrast, for environmental-robust downstream models, the primary of speech enhancement \cite{NR_ASR_2013Seltzer, NR_ASR_2022Hu, NR_E_SR_2019Zhou, Kim2023PASPA} is to prioritize regulating artificial noise levels. Consequently, the expected spectrum of these two model types is fundamentally different. Without awareness of the expected speech in the downstream network, the speech enhancement network faces confusion in deciding whether to minimize speech distortion or regulate artificial noise levels. The static network-based multi-task learning methods \cite{han2022dynamic} are less adaptable, making it hard to achieve universal speech enhancement.

Dynamic neural network, as surveyed by Han et al. \cite{han2022dynamic}, is a prominent topic known for their ability to adapt structures or parameters to different inputs, offering advantages in accuracy, computational efficiency, and adaptability. This adaptability makes the dynamic neural network have the potential to realize universal speech enhancement. The dynamic neural network has been widely used in computer vision \cite{wang2020glance,kirillov2020pointrend} and natural language processing \cite{elbayad2019depth,hansen2019neural}. However, it is rarely used for speech processing \cite{tavarone2018conditional}, especially for speech enhancement to the best of our knowledge. Relevant studies in this domain include the Multi-gate Mixture-of-Experts (MMOE) \cite{ma2018modeling} and task-specific weighting prediction approaches \cite{wang2019tafe, kang_incorporating_2020}. MMOE learns to model task relationships from data, but cannot used for unseen downstream models. The task-specific weighting prediction focuses on learning parameters specific to tasks. 
Therefore, despite the potential of dynamic neural networks to realize universal speech enhancement, the current landscape of dynamic neural networks does not align with the requirements for achieving universal speech enhancement.

In this paper, as depicted in Fig. \ref{fig:Plugin-SE concept} (b), we introduce a novel weight prediction approach that learns module relationships from downstream training information. Within the weight prediction module, downstream training information is systematically encoded, embedded, and mapped into the gate parameters. This conversion process enables the speech enhancement model to understand the expected speech signals, even when dealing with unseen downstream models.
We propose a novel dynamic speech enhancement framework, named Plugin Speech Enhancement (Plugin-SE), which integrates a common trunk speech enhancement module, a gate module, and a weight prediction module. The Plugin-SE is a kind of universal speech enhancement framework, which is flexible to most downstream network. The common trunk module is designed to minimize speech distortion and regulate artificial noise levels. The gate module, akin to the observation addition method used in robust Automatic Speech Recognition (ASR) \cite{iwamoto2022bad}, plays a crucial role. We observed that the decision of whether to use data augmentation techniques in downstream network training, particularly for noise injection, play a decisive role in determining the gate parameter. Therefore, the weight prediction is employed to transfer the downstream task and data augmentation method into the gate parameter.
This innovative approach empowers Plugin-SE to comprehend the target speech of the downstream model and seamlessly adapt to various tasks without requiring manual parameter adjustments.

This paper is organized as follows:
Section \ref{sec:Conventional Method} provides an overview of plugin speech enhancement, covering speech enhancement for single-task models, three types of loss functions for speech enhancement, and common trunk training.
Section \ref{sec:plugin SE} delves into the details of plugin speech enhancement, encompassing weight prediction training and the inference stages.
Section \ref{sec:Exp. results} presents the analysis of different downstream models, comparing the performance of plugin speech enhancement with baseline and fine-tuning methods.
Finally, Section \ref{sec:Conclusion} offers concluding remarks.

\section{Preliminary}
\label{sec:Conventional Method}

The plugin or universal speech enhancement represents a form of speech enhancement applicable to a broad spectrum of downstream speech processing tasks. During the inference stage, employing universal speech enhancement with a specific downstream model can be viewed as a environmental-robust single-task model. To provide a foundation for understanding the proposed plugin speech enhancement, we begin by introducing the environmental-robust single-task model and the pre-training method for speech enhancement.

\subsection{speech enhancement for single task model}
\label{ssec:Inference stage of environmental-robust system}

In Figure \ref{fig:Plugin-SE concept} (a), we present the single-task model \cite{iwamoto2022bad}. To simulate the inference stage, we randomly select the speech signal vector $\mathbf{s}$ and the noise signal vector $\mathbf{n}$ from the clean speech set and noise set, respectively. The input to the speech enhancement network is the noisy speech signal $\mathbf{x} = \mathbf{s} + \mathbf{n}$.
The enhanced speech for the downstream model, denoted as $\mathbf{s}_{mix}$, is the addition of the enhanced speech $\mathbf{\hat{s}}$ and the noisy speech $\mathbf{x}$. This can be expressed as:
\begin{equation}
\begin{aligned}
\mathbf{s}_{mix} &= (1-w)\mathbf{\hat{s}} + w\mathbf{x}\\
\mathbf{\hat{s}} &= f_{SE}\left(\mathbf{x}; \theta_{SE} \right)\\
\end{aligned}
\label{eq:enhanced speech}
\end{equation}
where $w$ is the gate parameter, and $\theta_{SE}$ represents the parameters of the speech enhancement network. The output of the downstream model, denoted as $\mathbf{\hat{v}_{x}}$, is given by
\begin{equation}
\mathbf{\hat{v}_{x}} = f_{DS}\left(\mathbf{\hat{s}}; \theta_{DS} \right)
\label{eq:downstream model}
\end{equation}
where $\theta_{SE}$ refers to the parameters of the downstream network.

\subsection{speech enhancement pre-training}
\label{ssec:SE pre-training}

There are three types of pre-training speech enhancement loss functions used for speech enhancement without downstream model, speech enhancement with a specified downstream model and speech enhancement with various downstream models, respectively.

1) For speech enhancement without downstream model, the parameters $\theta_{SE}$ of the SE model are updated by comparing the target signal $\mathbf{s}$ with the mapped signal $\mathbf{\hat{s}}$ generated by SE model. The Signal-to-Distortion Ratio (SI-SDR) loss \cite{le2019sdr} is commonly used for training the SE model:
\begin{equation}
\begin{aligned}
L_{\text{SI-SDR}}\left(\mathbf{\hat{s}}, \mathbf{s} \right) &= -\frac{10}{N}\sum_{i=1}^N \log_{10}\frac{||\mathbf{s}_t||^2}{||\mathbf{e}_d||^2}\\
\mathbf{s}_t &= \frac{\Braket{\mathbf{\hat{s}},\mathbf{s}}\mathbf{s}}{||\mathbf{s}||^2} \\
\mathbf{e}_d &= \mathbf{\hat{s}} - \mathbf{s}_t \\
\end{aligned}
\label{eq:SISNR}
\end{equation}

2) For speech enhancement with a specific downstream model, the parameters $\theta_{SE}$ and gate parameter $w$ are updated by comparing the target signal $\mathbf{s}$ with the enhanced signal $\mathbf{\hat{s}}$ and comparing the estimated features $\mathbf{\hat{v}}_x$ and $\mathbf{\hat{v}}_s$ with a trade-off parameter $\lambda_{CM}$ as follows:
\begin{equation}
L_{CM} = L_{\text{SI-SDR}}\left(\mathbf{\hat{s}}, \mathbf{s} \right)
        + \lambda_{CM}L_{DS}\left(\mathbf{\hat{v}}_x, \mathbf{\hat{v}}_s   \right)
\label{eq:CM_loss}
\end{equation}
where the estimated speech features $\mathbf{\hat{v}}_s$ are given by,
\begin{equation}
\mathbf{\hat{v}_s} = f_{DS}\left(\mathbf{s}; \theta_{DS} \right)
\label{eq:mapping_function}
\end{equation}
and the Kullback-Leibler (KL) Divergence $D_{KL}$ is utilized as downstream loss function $L_{DS}$ for ASR, SV, Hubert, etc.:
\begin{equation}
D_{KL}\left(\mathbf{\hat{v}}_x, \mathbf{\hat{v}}_s   \right) = \frac{1}{N}\sum{\mathbf{\hat{v}}_s \log{\frac{\mathbf{\hat{v}}_s}{\mathbf{\hat{v}}_x}}}
\label{eq:KL_divergence}
\end{equation}

3) For universal or common trunk speech enhancement, where the downstream model is not specified, the enhanced speech is based on reducing speech distortion and regulating artificial noise levels. The parameters of common trunk speech enhancement are updated by the following loss function $L_{ct}$,
\begin{equation}
L_{ct} = L_{\text{SI-SDR}}\left(\mathbf{\hat{s}}, \mathbf{s} \right)
        + \lambda_{ct}L_{\text{SI-SAR}}\left(\mathbf{\hat{s}}, \mathbf{s}, \mathbf{n} \right)
\label{eq:PI_loss}
\end{equation}
where term $\lambda_{ct}$ controls the influence of the SI-SAR loss, allowing for a balanced optimization process. The SI-SAR loss is shown as,
\begin{equation}
\begin{aligned}
L_{\text{SI-SAR}}\left(\mathbf{\hat{s}}, \mathbf{s}, \mathbf{n} \right) &= -\frac{10}{N}\sum_{i=1}^N \log_{10}\frac{||\mathbf{x}_t||^2}{||\mathbf{e}_{art}||^2} \\
\mathbf{x}_t &\approx \Braket{\mathbf{\hat{s}},\mathbf{s}}\mathbf{s}/||\mathbf{s}||^2 + \Braket{\mathbf{\hat{s}},\mathbf{n}}\mathbf{n}/||\mathbf{n}||^2 \\
\mathbf{e}_{art} &= \mathbf{\hat{s}} - \mathbf{x}_t \\
\end{aligned}
\label{eq:SISAR}
\end{equation}



\section{The plugin speech enhancement framework}
\label{sec:plugin SE}

As shown in Fig. \ref{fig:Plugin-SE concept} (b), the distinction between plugin speech enhancement with downstream models and the single-task model lies in the fact that the gate parameter is the output of the weight prediction. The weight prediction process enables the speech enhancement model to understand the features of the downstream model. The central challenge in universal speech enhancement shifts towards determining the type of information that should be utilized as inputs for the weight prediction.

\subsection{weight prediction}
\label{ssec:trade-off parameter training}

Downstream models exhibit numerous differences, encompassing aspects such as model structure, task, data augmentation techniques, and optimization methods. Analyzing the relationship among these models draws inspiration from discussions on artificial noise in ASR systems \cite{vincent2006performance, iwamoto2022bad}, particularly relevant to environmental-robust models.
Firstly, artificial noise remains unseen for environmental-robust downstream networks. Whether background noise is considered in the downstream network depends on whether it's added to the input speech signal during the network's training.
Secondly, for noise-sensitive models where all noise types are unseen, both artificial and background noise hold significance. Notably, artificial noise tends to have a more pronounced impact than background noise for environmental-robust models.
Lastly, the gate structure parameter shows a strong correlation with the source-to-artificial noise ratio and final performance \cite{iwamoto2022bad}. 
Therefore, the decision of whether to use the noise injection method emerges as a pivotal factor for the gate structure parameters. This implies a nonlinear mapping from noise injection information to gate structure parameters. Based on our experimental results, we find that task and the decision of whether to use data augmentation serve as weight prediction inputs.

The weight prediction structure is a neural network with information encoding, embedding and parameter mapping. The task and whether to use data augmentation are first encoded as task identification (ID) and $\{ 0,1 \}$. The task ID is then transformed into an embedding space $E_{task}$, represented as $E_{task} = \mathbf{M}[id]$, where $\mathbf{M}$ is an embedding matrix and $id$ is the index of tasks. Subsequently, the features, concatenating whether does noise inference $B_{NI} \in \{0, 1\}$ and embedded task, are mapped into the gate parameter $\hat{w}$ by a DNN structure:

\begin{equation}
\hat{w} = f_{map}(E_{task}, B_{NI}; \theta_{map})
\end{equation}

The target gate parameter $w \in [0,1]$ is the optimized parameter based on the downstream loss function:
\begin{equation}
L_w=L_{DS}\left(\mathbf{\hat{v}_x}, \mathbf{\hat{v}}_s \right)
\label{eq:adapter training}
\end{equation}
where $\mathbf{\hat{v}_x} = f_{DS}\left(\mathbf{s}_{mix}; \theta_{DS} \right)$, $\mathbf{\hat{v}_s} = f_{DS}\left(\mathbf{s}; \theta_{DS} \right)$ and $\mathbf{s}_{mix} = (1-w)\mathbf{\hat{s}} + w\mathbf{x}$. The KL divergence is used as the downstream loss function.

The embedding and parameter mapping are optimized by the mean square error loss function.

\subsection{Inference stage}
\label{ssec:Inference and test stage}

\begin{figure}[tb]
    \centering
    \centerline{\includegraphics[width=10cm]{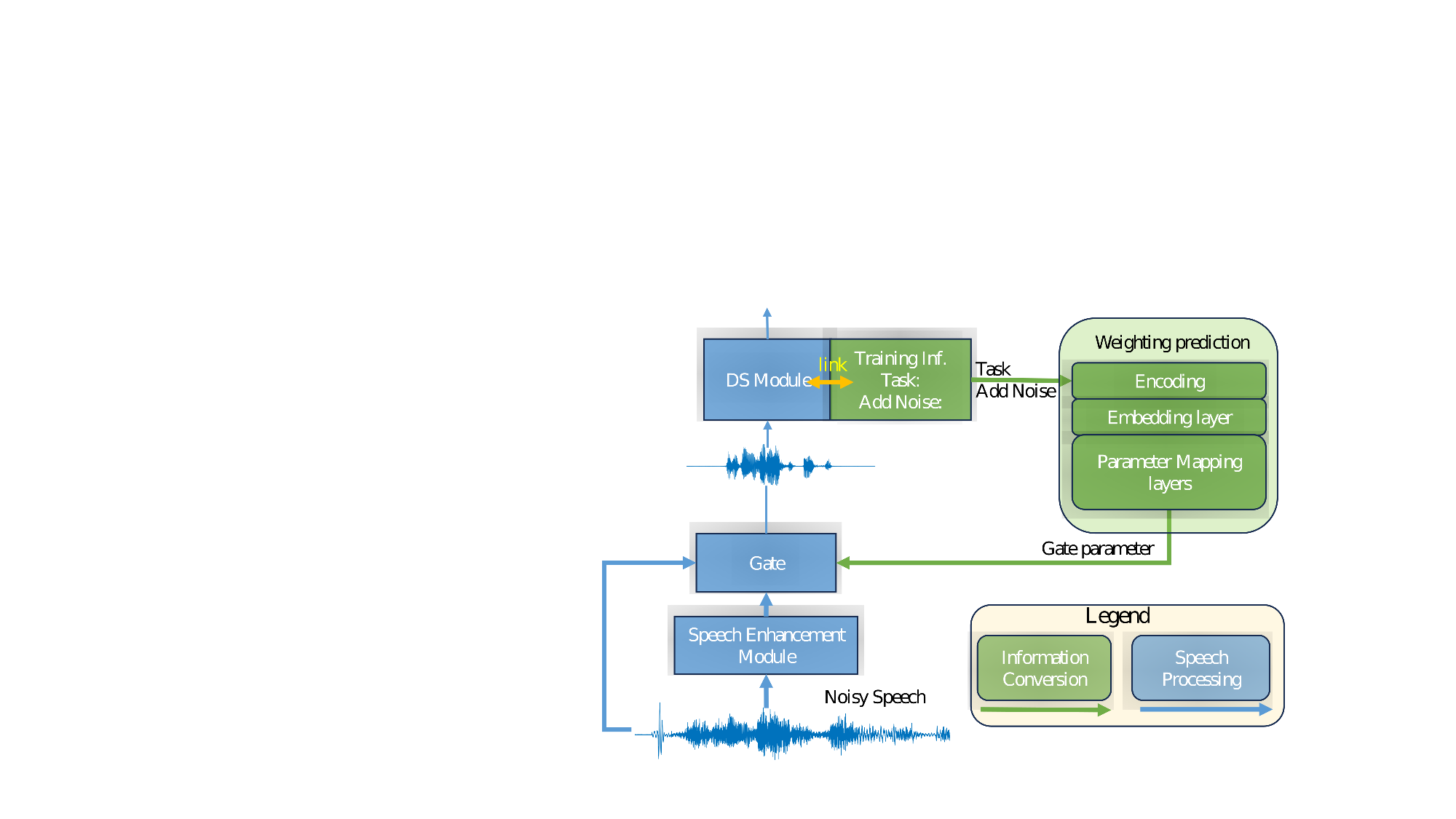}}
\caption{The inference stage of the plugin speech enhancement for single task. The gate parameter is determined before the speech processing.}
\label{fig:plugin_test}
\end{figure}

Fig. \ref{fig:Plugin-SE concept} (b) illustrates the inference stage of the proposed Plugin-SE framework. For specific downstream model, the inference for a single task is shown in Fig. \ref{fig:plugin_test}. The speech enhancement parameters $\theta_{SE}$ and downstream model $\theta_{DS}$ are initially provided. Subsequently, the task ID and noise inference $B_{NI}$ of downstream model are embedded and represented in the gate parameter $\hat{w}$. Finally, the inference result is obtained by the deployed model, i.e.,
\begin{equation}
\begin{aligned}
\mathbf{\hat{v}_x} &= f_{DS}\left(\mathbf{s}_{mix}; \theta_{DS} \right)\\
\mathbf{s}_{mix}   &= (1-\hat{w})\mathbf{\hat{s}} + \hat{w}\mathbf{x}\\
\mathbf{\hat{s}}   &= f_{SE}\left(\mathbf{x}; \theta_{SE} \right) \\
\hat{w}            &= f_{map}(E_{task}, B_{NI}; \theta_{map}) \\
\end{aligned}
\end{equation}

\section{Experiments}
\label{sec:Exp. results}
\subsection{Experimental settings}
\label{ssec:Experimental settings}

In this experiment, the Plugin-SE encompasses four tasks: speech enhancement, SV, ASR, and Hubert. The dimension of matrix $\mathbf{M}$ of weight prediction network is 10 x 4. We utilize fully-connected neural networks as mapping layers, comprising three hidden layers with ReLU activation functions, each consisting of 256 features. Sigmoid activation function is applied after the output layer.

Following pre-trained models are utilized: HapNet \cite{wang2023harmonic} for speech enhancement, Conformer \cite{burchi2021efficient} for ASR, ResNet34 \cite{Chung2018VoxCeleb2DS} and ECAPA-TDNN \cite{desplanques20_interspeech} for SV, and Hubert \cite{hsu2021hubert} for speech feature representation. While the training data of ECAPA-TDNN is sourced from VoxCeleb2, the remaining models utilize the original training data.
For fine-tuning HapNet, we utilize the DNS-challenge database \cite{dubey2022icassp} with random SNR ranges from 0 to 20 dB. Half of the utterances are subjected to convolution with both random synthetic and real room impulse responses (RIRs) prior to mixing.
The common trunk SE model, in this experiment, is called fine-tuned HapNet, which is optimized based on Eq. \ref{eq:PI_loss}. 

We conducted a comparative analysis involving the pre-trained HapNet, HapNet common trunk model, HapNets for ASR, SV and Hubert with the Plugin-SE methods, as well as fine-tuning HapNets for ASR, SV and Hubert as discussed in Section \ref{sec:Conventional Method}. These are referred to as HapNet, HapNet-CT, HapNet-PlugASR, HapNet-PlugSV, HapNet-PlugHu, HapNet-FT-ASR, HapNet-FT-SV and HapNet-FT-Hu, respectively. ResNet34 and ECAPA-TDNN are pre-trained environmental-robust SV models, and ResNet34 without data augmentation (ResNet34 w/o) is noise-sensitive SV model for testing the model-insensitivity of the proposed Plugin-SE. The relationship pairs with the estimated parameter $\hat{w}$  of Plugin-SE are shown in Table \ref{tab:plugin model}.

\begin{table}[!t]
\centering
\caption{The plugin speech enhancement model and relationship information table.}
\label{tab:plugin model}
\tabcolsep=0.15cm
\renewcommand\arraystretch {1}
\resizebox{11cm}{!}{
\begin{tblr}{c|ccc|c}
\hline \hline
\SetCell[r=2]{c}{SE}        & \SetCell[c=3]{c}{weight prediction} & &         & \SetCell[r=2]{c}{downstream network}                           \\
                            & task                  & noise injection & $\hat{w}$ &                                                               \\ \hline
\SetCell[r=6]{c}{HapNet-CT} & ASR                   & Yes             & 0.9       & conformer \cite{burchi2021efficient}         \\ \cline[dotted]{2-6} 
                            &  \SetCell[r=3]{c}{SV} & Yes             & 0.56      & ResNet34 \cite{Chung2018VoxCeleb2DS}         \\
                            &                       & Yes             & 0.56      & ECAPA-TDNN \cite{desplanques20_interspeech} \\
                            &                       & No              & 0.02      & ResNet34 w/o                                                  \\ \cline[dotted]{2-6}
                            & SE                    & No              & 0         & None                                                          \\ \cline[dotted]{2-6}
                            & Representation        & No              & 0.01      & Hubert \cite{hsu2021hubert}                  \\
 \hline \hline
\end{tblr}}
\end{table}

In our experiments, the trade-off parameters $\lambda_{CM}$ and $\lambda_{PI}$ are set to 0.01. For network training, we utilize the Adam optimization algorithm \cite{kingma2014adam} with an initial learning rate of $10^{-5}$, coupled with exponential decay parameterized by $\gamma = 0.9$ for learning rate adjustment. 

In our experiments, we evaluate the SE, ASR, SV and Representation tasks using the DNS challenge test set, AIshell2 test set, VoxCeleb1 test set, respectively. The noisy speech used for testing is either originally from a noisy corpus or generated through mixing speech and noise at six types of the SNR levels (e.g., -5 dB, 0 dB, 5 dB, 10 dB, 15 dB and 30 dB). We compare the performance of the different models based on the Perceptual Evaluation of Speech Quality (PESQ) \cite{recommendation2001perceptual}, Short-Time Objective Intelligibility (STOI) \cite{taal2011algorithm}, as well as SIG, BAK, and OVL metrics \cite{reddy2021dnsmos} for the SE task. For the SV task, we employ the Equal Error Rate (EER) and the minimum value of the Detection Cost Function (minDCF), while for the ASR task, we use the Word Error Rate (WER). For the Representation task, we use the Unit Error Rate (UER), Mean Square Error (MSE), and KL Divergence (KLD). The ground truth for SE, ASR, and SV tasks is derived from labeled data, and the ground truth for the Representation task is determined based on the evaluated representation $\mathbf{\hat{v}}_s$ from clean speech $\mathbf{s}$.




\begin{figure}[tb]
    \centering
    \centerline{\includegraphics[width=13cm]{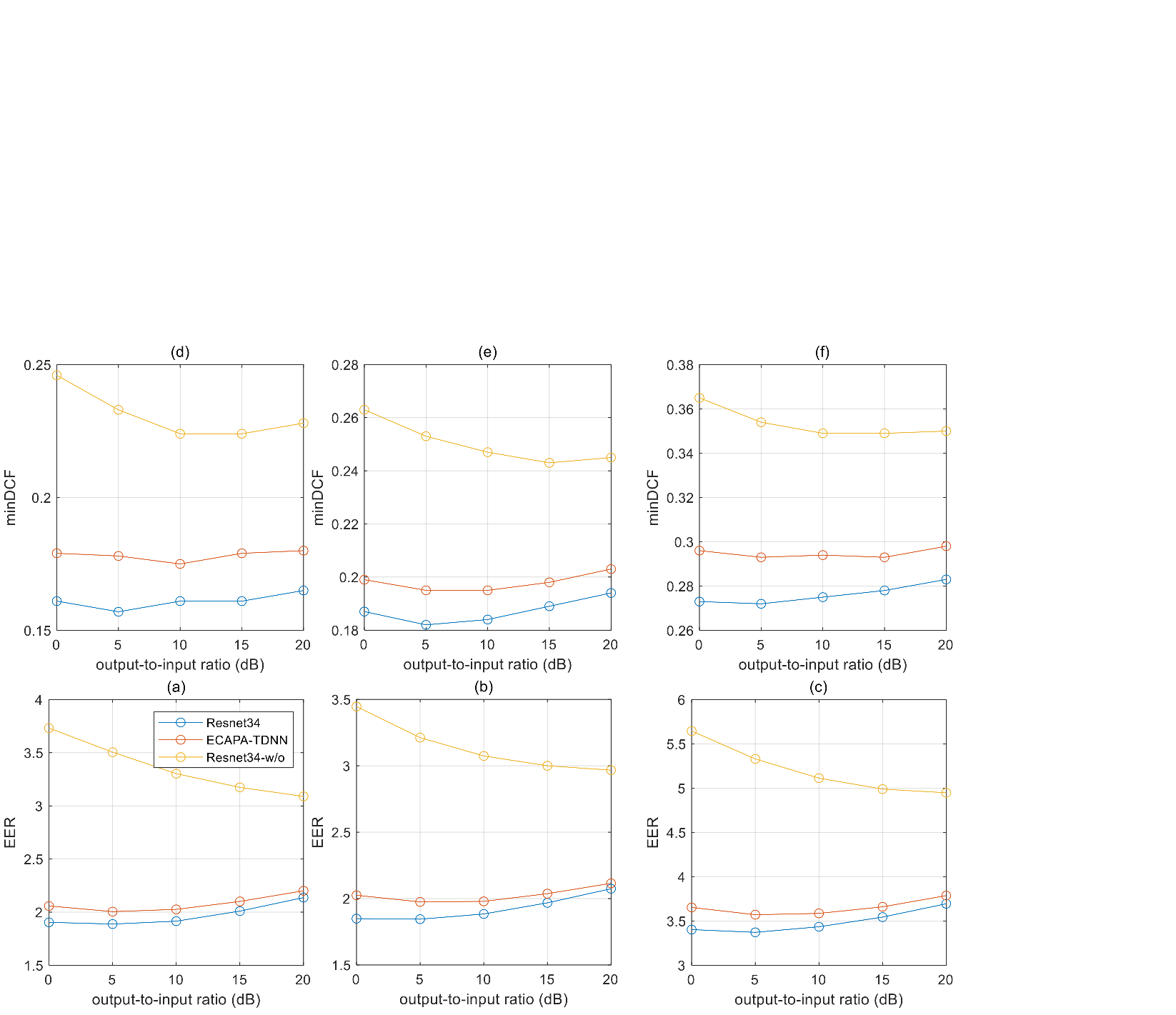}}
\caption{The relationship between output-to-input ratio and SV scores.}
\label{fig:OIR-EER}
\end{figure}

\begin{table}[!t]
\centering
\caption{The downstream-task scores for methods on noisy speech. (\textbf{Bold} indicates the best results, DCF indicates minDCF in this table, KL divergence is enlarged by $10^5$)}
\label{tab:results-noisy}
\tabcolsep=0.15cm
\renewcommand\arraystretch {1}
\resizebox{12cm}{!}{
\begin{tabular}{l|c|cccccc|ccc}
\hline \hline
Task      & \multicolumn{1}{c|}{ASR} & \multicolumn{6}{c|}{SV}                                                                                                                 & \multicolumn{3}{c}{Hubert}                                         \\ \hline
database  & \multicolumn{1}{c|}{}    & \multicolumn{2}{c|}{vox1\_O}                       & \multicolumn{2}{c|}{vox1\_E}                       & \multicolumn{2}{c|}{vox1\_H}  & \multicolumn{1}{c}{} & \multicolumn{1}{c}{} & \multicolumn{1}{c}{} \\
Metric    & WER                      & EER           & \multicolumn{1}{l|}{DCF}           & EER           & \multicolumn{1}{l|}{DCF}           & EER           & DCF           & UER                  & MSE                  & KLD                   \\ \hline
None      & 7.45                     & 2.10          & \multicolumn{1}{l|}{0.17}          & 2.12          & \multicolumn{1}{l|}{0.20}          & 3.81          & 0.29          & 0.70                 & 12.63                & 0.87                 \\
HapNet    & 8.51                     & 2.46          & \multicolumn{1}{l|}{0.19}          & 2.50          & \multicolumn{1}{l|}{0.22}          & 4.35          & 0.31          & 0.40                 & 6.85                 & 0.27                 \\
-CT       & 7.80                     & 2.30          & \multicolumn{1}{l|}{0.18}          & 2.33          & \multicolumn{1}{l|}{0.21}          & 4.11          & 0.30          & 0.40                 & 6.69                 & 0.25                 \\ \hline
-PlugSE-ASR & \textbf{6.40}          & 1.75          & \multicolumn{1}{l|}{\textbf{0.14}} & 1.89          & \multicolumn{1}{l|}{\textbf{0.18}} & 3.36          & \textbf{0.27} & 0.55                 & 10.72                 & 0.64                 \\
-FT-ASR   & 6.74                     & 3.05          & \multicolumn{1}{l|}{0.19}          & 2.91          & \multicolumn{1}{l|}{0.23}          & 4.87          & 0.31          & 0.47                 & 6.91                 & 0.26                 \\ \hline
-PlugSE-SV & 6.60                    & \textbf{1.69} & \multicolumn{1}{l|}{0.15}          & \textbf{1.75} & \multicolumn{1}{l|}{\textbf{0.18}} & \textbf{3.23} & \textbf{0.27} & 0.53                 & 10.17                 & 0.58                 \\
-FT-SV    & 8.03                     & 2.40          & \multicolumn{1}{l|}{0.19}          & 2.34          & \multicolumn{1}{l|}{0.21}          & 4.14          & 0.30          & 0.45                 & 6.43        & \textbf{0.24}                 \\ \hline
-PlugSE-Hu & 7.94                    & 2.46          & \multicolumn{1}{l|}{0.18}          & 2.42          & \multicolumn{1}{l|}{0.20}          & 4.13          & 0.29          & 0.40                 & 6.82                 & 0.26                 \\
-FT-Hu    & 8.40                     & 2.64          & \multicolumn{1}{l|}{0.18}          & 2.60          & \multicolumn{1}{l|}{0.21}          & 4.37          & 0.30          & \textbf{0.38}        & \textbf{6.41}                 & \textbf{0.24}       
\\ \hline \hline
\end{tabular}
}
\end{table}

\begin{table}[!t]
\centering
\caption{The different task scores for methods on clean speech. (\textbf{Bold} indicates the best results, and DCF indicates minDCF in this table)}
\label{tab:results-clean}
\tabcolsep=0.2cm
\renewcommand\arraystretch {1}
\resizebox{10cm}{!}{
\begin{tabular}{l|c|cccccc}
\hline \hline
Task        & \multicolumn{1}{c|}{ASR} & \multicolumn{6}{c}{SV}                                                                                                                  \\ \hline
database    & \multicolumn{1}{c|}{}    & \multicolumn{2}{c|}{vox1\_O}                       & \multicolumn{2}{c|}{vox1\_E}                       & \multicolumn{2}{c}{vox1\_H}   \\
Metric      & WER                      & EER           & \multicolumn{1}{l|}{DCF}           & EER           & \multicolumn{1}{l|}{DCF}           & EER           & DCF           \\ \hline
None        & 3.48                     & \textbf{0.96} & \multicolumn{1}{l|}{\textbf{0.09}} & \textbf{1.12} & \multicolumn{1}{l|}{\textbf{0.13}} & \textbf{1.12} & \textbf{0.20} \\
HapNet      & 4.06                     & 1.06          & \multicolumn{1}{l|}{0.10}          & 1.26          & \multicolumn{1}{l|}{0.14}          & 1.26          & 0.21          \\
-CT         & 3.50                     & 1.01          & \multicolumn{1}{l|}{0.10}          & 1.16          & \multicolumn{1}{l|}{\textbf{0.13}} & 1.16          & \textbf{0.20} \\ \hline
-PlugSE-ASR & \textbf{3.46}            & 0.99          & \multicolumn{1}{l|}{0.10}          & 1.13          & \multicolumn{1}{l|}{\textbf{0.13}} & 2.10          & \textbf{0.20} \\
-FT-ASR     & 3.49                     & 1.10          & \multicolumn{1}{l|}{0.10}          & 1.30          & \multicolumn{1}{l|}{0.14}          & 2.36          & 0.21          \\ \hline
-PlugSE-SV  & 3.45                     & \textbf{0.96} & \multicolumn{1}{l|}{\textbf{0.09}} & 1.13          & \multicolumn{1}{l|}{\textbf{0.13}} & 1.13          & \textbf{0.20} \\
-FT-SV      & 3.54                     & 1.14          & \multicolumn{1}{l|}{0.11}          & 1.26          & \multicolumn{1}{l|}{0.14}          & 1.26          & 0.21         
\\ \hline \hline
\end{tabular}
}
\end{table}

\subsection{Experiment 1: Evaluation with gate structure}
\label{ssec:Exp:oa parameter}

In addition to ResNet34, and ResNet34 w/o, we also train the ECAPA-TDNN using the same training set as pre-trained ResNet34. We use five types of SNR levels (e.g., 0 dB, 5 dB, 10 dB, 15 dB and 20 dB) to illustrate the relationship between the enhanced-to-mix ratio and SV scores based on different data augmentation and network structures.

\begin{table}[!t]
\centering
\caption{The SE scores for methods on noisy speech. (\textbf{Bold} indicates the best results)}
\label{tab:results-SE}
\tabcolsep=0.2cm
\renewcommand\arraystretch {1}
\resizebox{7cm}{!}{
\begin{tabular}{l|ccccc}
\hline \hline
Task      & \multicolumn{5}{c}{SE}                                                        \\
Metric    & OVRL          & SIG           & BAK           & pesq          & stoi          \\ \hline
None      & 2.05          & 2.90          & 2.12          & 1.26          & 0.83          \\
HapNet    & \textbf{3.30} & \textbf{3.54} & \textbf{4.15} & \textbf{2.79} & \textbf{0.94} \\
-CT       & 3.24          & 3.52          & 4.06          & 2.77          & \textbf{0.94} \\ \hline
-Plug-ASR   & 2.42          & 3.37          & 2.53          & 1.47          & 0.88          \\
-FT-ASR & 3.24          & 3.51          & 4.09          & 2.45          & 0.93          \\ \hline
-Plug-SV    & 2.54          & 3.44          & 2.71          & 1.56          & 0.89          \\
-FT-SV  & 3.24          & 3.51          & 4.07          & 2.73          & \textbf{0.94} \\ \hline
-Plug-Hu    & 3.22          & 3.53          & 4.02          & 2.72          & \textbf{0.94} \\
-FT-Hu  & 3.26          & 3.53          & 4.09          & 2.73          & \textbf{0.94}\\ \hline \hline
\end{tabular}
}
\end{table}

\begin{table}[!t]
\centering
\caption{The SV scores for methods on noisy speech. (\textbf{Bold} indicates the best results)}
\label{tab:results-SV}
\tabcolsep=0.2cm
\renewcommand\arraystretch {1}
\resizebox{8cm}{!}{
\begin{tabular}{l|cc|cc|cc}
\hline \hline
              & \multicolumn{2}{c|}{vox1\_O}  & \multicolumn{2}{c|}{vox1\_E}  & \multicolumn{2}{c}{vox1\_H}   \\
              & EER           & DCF           & EER           & DCF           & EER           & DCF           \\ \hline
None          & 2.10          & 0.17          & 2.12          & 0.20          & 3.81          & 0.29          \\
HapNet        & 2.46          & 0.19          & 2.50          & 0.22          & 4.35          & 0.31          \\
-FT-SV        & 2.40          & 0.19          & 2.34          & 0.21          & 4.14          & 0.30          \\
-Plugin-SV    & 1.69          & 0.15          & 1.75          & 0.18          & 3.23          & 0.27          \\
-Plugin-SV-FT & \textbf{1.58} & \textbf{0.13} & \textbf{1.61} & \textbf{0.16} & \textbf{2.87} & \textbf{0.25} \\ \hline \hline
\end{tabular}
}
\end{table}

\begin{figure}[t]
    \centering
    \centerline{\includegraphics[width=12cm]{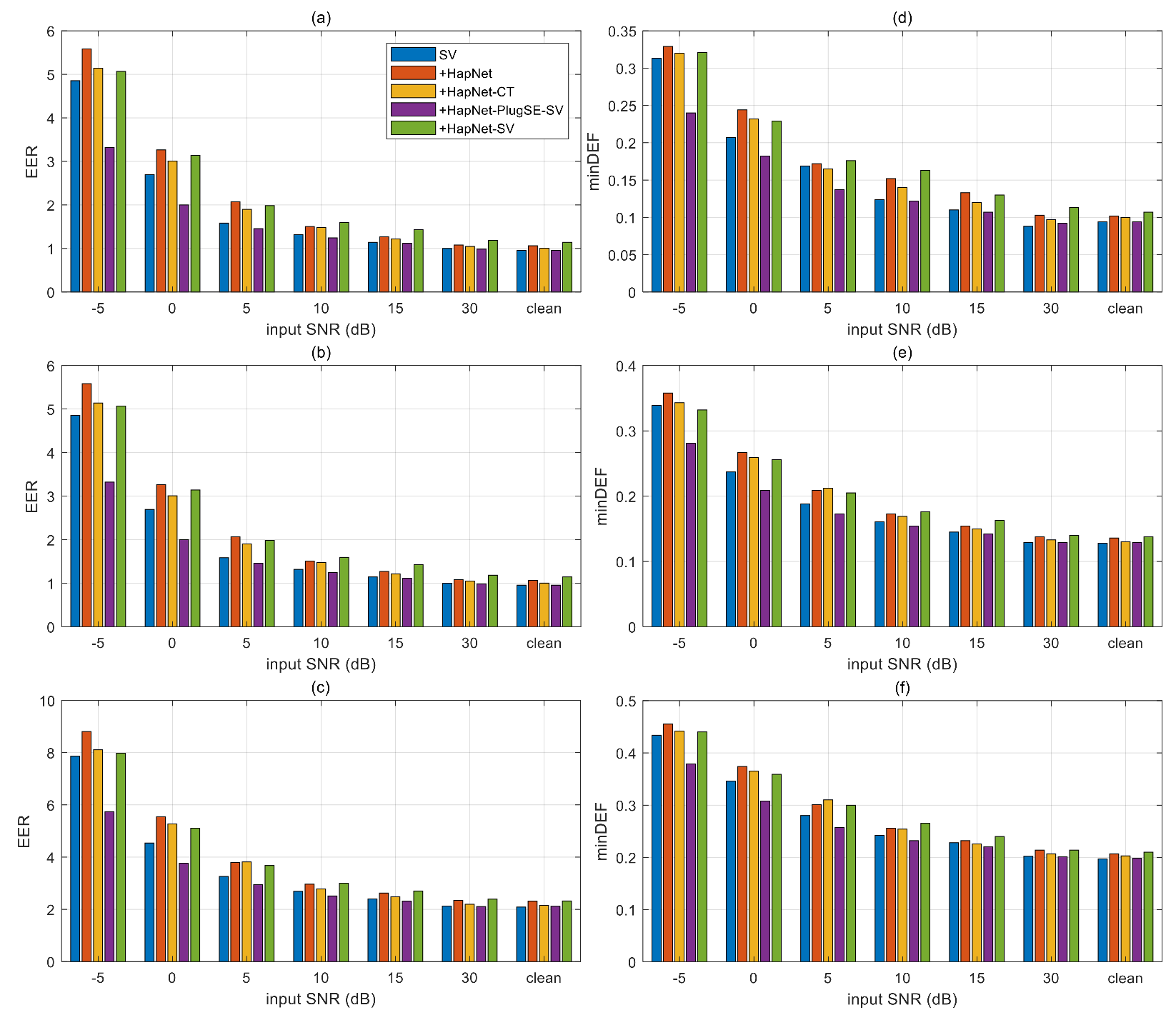}}
\caption{The SV scores for methods on different test datasets. (vox1\_O\_cleaned is used for (a) and (d), vox1\_E\_cleaned is used for (b) and (e), and vox1\_H\_cleaned is used for (c) and (f).) }
\label{fig:details SV}
\end{figure}

\begin{figure}[t]
    \centering
    \centerline{\includegraphics[width=12cm]{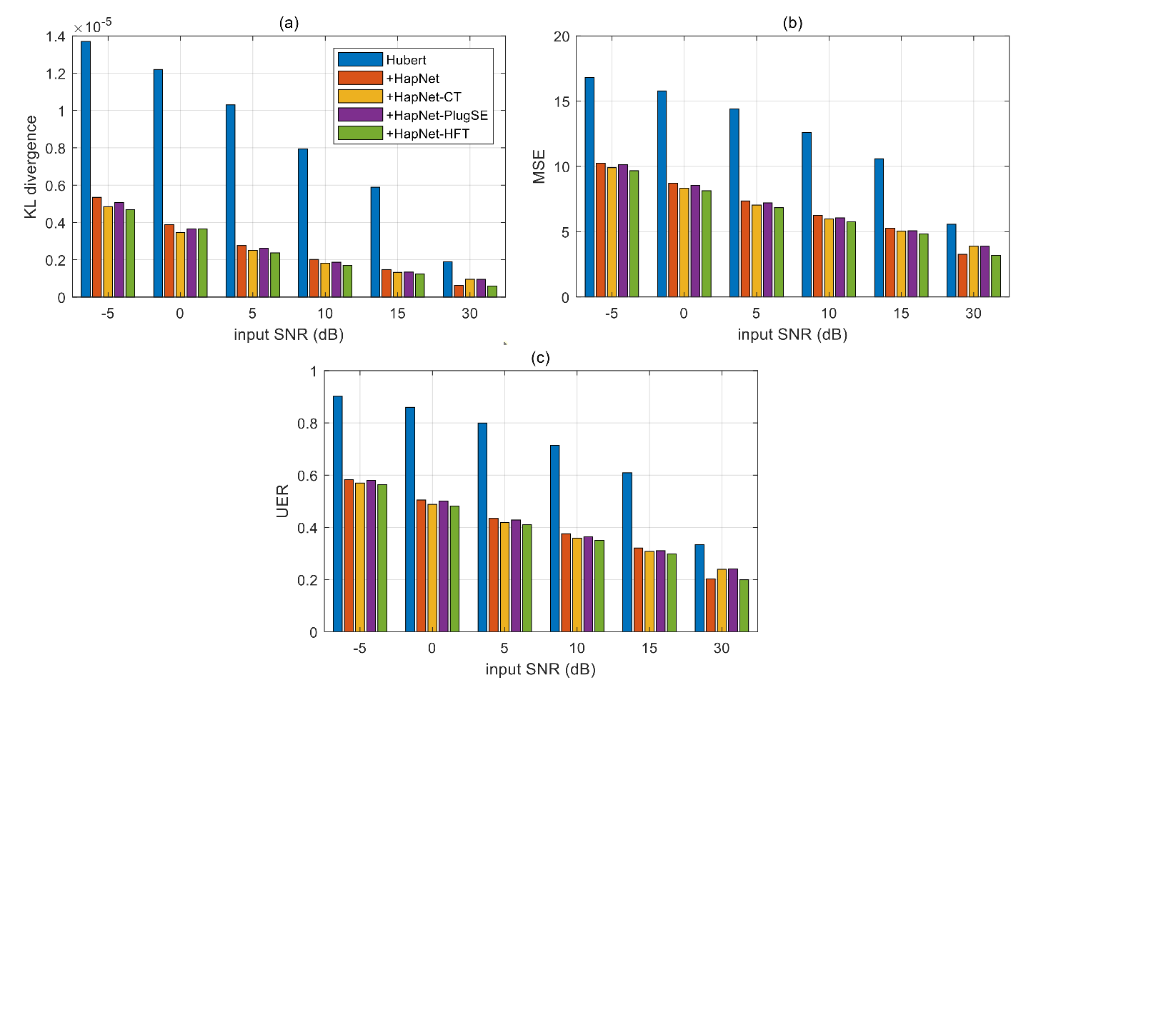}}
\caption{The Hubert scores for methods on noisy speech.}
\label{fig:details Hubert}
\end{figure}

\begin{figure}[t]
    \centering
    \centerline{\includegraphics[width=12cm]{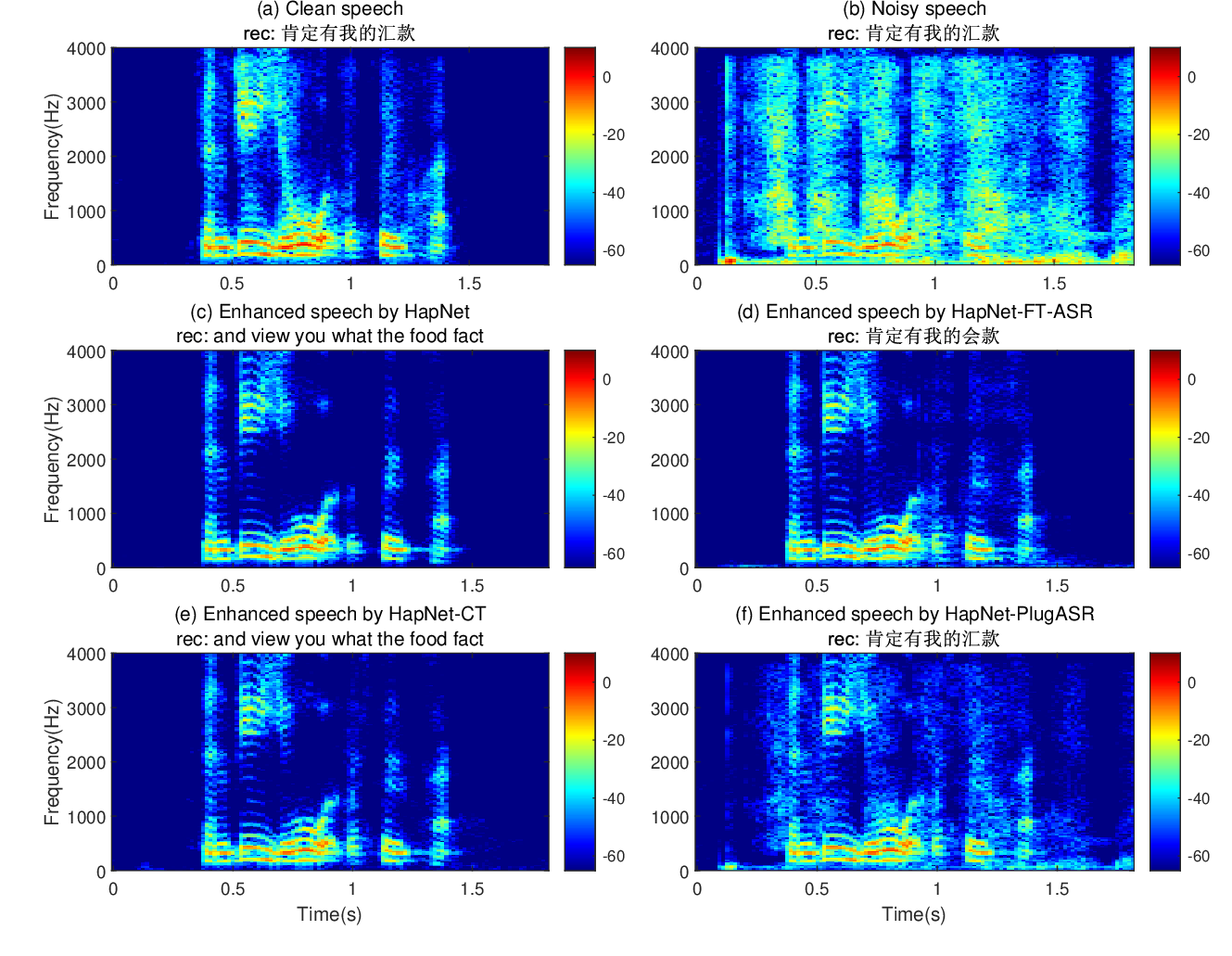}}
\caption{Spectrogram comparison of clean speech, noisy speech,
HapNet, HapNet-FT-ASR, HapNet-CT and HapNet-PlugASR from 0 to 4 kHz.}
\label{fig:spectrogram}
\end{figure}

Figure \ref{fig:OIR-EER} illustrates the relationship between the average output-to-input ratio (OIR) $r$ and SV scores, where 'output' and 'input' indicate the enhanced speech and noisy speech signal, and the OIR is highly related to the gate parameter $w$. The score curves of environmental-robust models, such as ResNet34 and ECAPA-TDNN, are similar, while the curves differ between ResNet34 and ResNet34-w/o.
This observation highlights that the optimal gate parameter is strongly influenced by the downstream data augmentation method, and to a lesser extent by the downstream structure. Based on this analysis, unifying the speech enhancement model with the static neural network proves challenging. From this experiment, whether using noise injection and the task are main additional information for Plugin-SE.
    
Figure \ref{fig:OIR-EER} also reveals that the parameter $r$ is typically higher for noise-sensitive models compared to environmental-robust models. This result is further validated in the subsequent experiment.

\subsection{Experiment 2: Plugin speech enhancement performances}
\label{ssec:Exp3:PSE tests}

Table \ref{tab:results-noisy} and Table \ref{tab:results-clean} present the scores for downstream tasks using the mentioned methods on noisy speech test data and clean speech test data, respectively.
During the noisy speech test, the plugin speech enhancement exhibits the best performance in ASR and SV tests, achieved with well-trained parameters $\hat{w} = 0.9 $ for ASR and $\hat{w} = 0.56 $ for SV, respectively.
In the case of clean speech tests, the plugin speech enhancement demonstrates comparable performance to None speech enhancement model. This phenomenon may be attributed to the fact that the speech enhancement model can mitigate artificial noise, as indicated by the common trunk loss function, i.e., Eq. \ref{eq:PI_loss}.

In the Hubert test, the fine-tuning method performed the best, followed by the HapNet-CT, which can be considered as the HapNet-PlugSE-Hu with parameter $\hat{w} = 0 $. The plugin speech enhancement, with well-trained parameter $\hat{w} = 0.01$, ranked third. Two reasons may contribute to this phenomenon. First, the Hubert training is based on self-supervised methods, which may make Hubert sensitive to almost noise. Second, similar to the result of Experiment 1, as the parameter $w$ decreases, some performances are better for the noise-sensitive model.

Table \ref{tab:results-SE} shows the performances of mentioned methods for speech enhancement task. The best-performing method is pre-trained HapNet, followed by the common trunk HapNet. The plugin speech enhancement methods for ASR and SV perform noticeably worse in speech enhancement tasks. This can be attributed to the error parameter $w$. In the context of speech enhancement task, the downstream model, which can be likened to an Identity matrix, proves to be sensitive to all forms of noise. It is more appropriate to define the parameter $w$ as 0. The fine-tuning method also exhibits satisfactory performance in the speech enhancement task due to the utilization of the loss function in equation \ref{eq:CM_loss}.

Table \ref{tab:results-SV} shows the performances of the Plugin-SE compared with the downstream fine-tuned model. The fine-tuned model is trained by frozen Plugin-SE model, using same training database and same loss function, referred to as HapNet-Plugin-SV-FT. This fine-tuned model learns the artificial noise, so that the background noise and artificial noise are both seen, which can give better performance. The proposed Plugin-SE performance is also acceptable, indicating that the Plugin-SE can be directly used for downstream models.

In detail, Fig. \ref{fig:details SV} shows the performance of SV task in different input SNR levels, where Fig. \ref{fig:details SV}(a) and Fig. \ref{fig:details SV}(d) are scores of vox1\_O\_cleaned, Fig. \ref{fig:details SV}(b) and Fig. \ref{fig:details SV}(e) are scores of vox1\_E\_cleaned, Fig. \ref{fig:details SV}(c) and Fig. \ref{fig:details SV}(f) are scores of vox1\_H\_cleaned. We could see the Plugin SE is better than other enhancement methods and None speech enhancement method in a noisy environment, especially for lower input SNR, and the Plugin SE is comparable to None enhanced processing in a quiet environment.
Fig. \ref{fig:details Hubert} shows the performance of Hubert test in different input SNR levels. The fine-tuning method is the best one, the common trunk model is comparable with SNR level smaller than 15 dB. However, when the input SNR is large enough, the performance of HapNet is better, which may be caused by the 0 dB to 20 dB SNR levels used for training the common trunk model.

\subsection{Spectrogram analysis}
\label{ssec:spectrogram analysis}

Fig. \ref{fig:spectrogram} provides a spectrogram comparison of the mentioned methods in the frequency range from 0 to 4 kHz, focusing primarily on the harmonic components. In this example, the ASR system can recognize the clean speech, noisy speech, enhanced speech by HapNet-FT-ASR and HapNet-PlugASR. In the case of HapNet-FT-ASR, a word is unrecognized but the pronounciation is the same. The HapNet and HapNet-CT transcribe the Chinese into English, maybe because some unvoiced structures are broken. Additionally, HapNet achieves highest PESQ performance. This test shows that the background noise primarily interferes with speech enhancement, while it has a minor impact on ASR model. However, some structural muffling may mislead the ASR system, and this error is typically generated during the enhancing process.

\section{Conclusion}
\label{sec:Conclusion}

In this study, we introduced the plugin speech enhancement framework, a novel dynamic neural network, to adapt to various downstream processing by incorporating the training information of downstream models into the gate parameter. This enables the speech enhancement model to understand the expected speech of downstream networks. The results demonstrate the effectiveness of our plugin approach in direct application to various speech processing tasks. Moving forward, enhancing the diversity of downstream models can improve the generalization capability of Plugin-SE, making it more universal.



\bibliographystyle{elsarticle-num} 
\bibliography{IEEEabrv,myabrv,refs}

\begin{thebibliography}{10}
\expandafter\ifx\csname url\endcsname\relax
  \def\url#1{\texttt{#1}}\fi
\expandafter\ifx\csname urlprefix\endcsname\relax\def\urlprefix{URL }\fi
\expandafter\ifx\csname href\endcsname\relax
  \def\href#1#2{#2} \def\path#1{#1}\fi

\bibitem{Loizou2013speech}
P.~C. Loizou, Speech Enhancement: Theory and Practice, 2nd Edition, CRC Press, Inc., USA, 2013.

\bibitem{telecom_2023moore}
B.~C. Moore, Speech processing for the hearing-impaired: successes, failures, and implications for speech mechanisms, Speech communication 41~(1) (2003) 81--91.

\bibitem{wang2022hgcn}
T.~Wang, W.~Zhu, et~al., Hgcn: Harmonic gated compensation network for speech enhancement, in: Proc.\ IEEE Int.\ Conf.\ Acoust., Speech, Signal Process., IEEE, 2022, pp. 371--375.

\bibitem{cui2023semi}
Z.~Cui, S.~Zhang, Y.~Chen, Y.~Gao, C.~Deng, J.~Feng, Semi-supervised speech enhancement based on speech purity, in: ICASSP 2023-2023 IEEE International Conference on Acoustics, Speech and Signal Processing (ICASSP), IEEE, 2023, pp. 1--5.

\bibitem{NR_ASR_2013Seltzer}
M.~L. Seltzer, D.~Yu, Y.~Wang, An investigation of deep neural networks for noise robust speech recognition, in: Proc.\ IEEE Int.\ Conf.\ Acoust., Speech, Signal Process., IEEE, 2013, pp. 7398--7402.

\bibitem{NR_ASR_2022Hu}
Y.~Hu, N.~Hou, et~al., Interactive feature fusion for end-to-end noise-robust speech recognition, in: Proc.\ IEEE Int.\ Conf.\ Acoust., Speech, Signal Process., IEEE, 2022, pp. 6292--6296.

\bibitem{chang2022end}
X.~Chang, T.~Maekaku, et~al., End-to-end integration of speech recognition, speech enhancement, and self-supervised learning representation, arXiv preprint arXiv:2204.00540 (2022).

\bibitem{NR_E_SR_2019Zhou}
J.~Zhou, T.~Jiang, et~al., Training multi-task adversarial network for extracting noise-robust speaker embedding, in: Proc.\ IEEE Int.\ Conf.\ Acoust., Speech, Signal Process., IEEE, 2019, pp. 6196--6200.

\bibitem{Kim2023PASPA}
W.~Kim, H.~Shin, et~al., \href{https://api.semanticscholar.org/CorpusID:259991578}{Pas: Partial additive speech data augmentation method for noise robust speaker verification}, ArXiv abs/2307.10628 (2023).
\newline\urlprefix\url{https://api.semanticscholar.org/CorpusID:259991578}

\bibitem{Shon2019VoiceIDLS}
S.~Shon, H.~Tang, J.~R. Glass, \href{https://api.semanticscholar.org/CorpusID:102352003}{Voiceid loss: Speech enhancement for speaker verification}, in: Interspeech, 2019.
\newline\urlprefix\url{https://api.semanticscholar.org/CorpusID:102352003}

\bibitem{NR_ER_2019Trian}
A.~Triantafyllopoulos, G.~Keren, et~al., Towards robust speech emotion recognition using deep residual networks for speech enhancement, in Interspeech (2019).

\bibitem{latif2023sparks}
S.~Latif, M.~Shoukat, F.~Shamshad, M.~Usama, Y.~Ren, H.~Cuayáhuitl, W.~Wang, X.~Zhang, R.~Togneri, E.~Cambria, B.~W. Schuller, Sparks of large audio models: A survey and outlook (2023).
\newblock \href {http://arxiv.org/abs/2308.12792} {\path{arXiv:2308.12792}}.

\bibitem{gslm}
K.~Lakhotia, E.~Kharitonov, et~al., On generative spoken language modeling from raw audio, Transactions of the Association for Computational Linguistics 9 (2021) 1336--1354.

\bibitem{speechlm}
Z.~Zhang, S.~Chen, et~al., Speechlm: Enhanced speech pre-training with unpaired textual data, arXiv preprint arXiv:2209.15329 (2022).

\bibitem{yang2023uniaudio}
D.~Yang, J.~Tian, X.~Tan, R.~Huang, S.~Liu, X.~Chang, J.~Shi, S.~Zhao, J.~Bian, X.~Wu, et~al., Uniaudio: An audio foundation model toward universal audio generation, arXiv preprint arXiv:2310.00704 (2023).

\bibitem{trinh2022unsupervised}
V.~A. Trinh, S.~Braun, Unsupervised speech enhancement with speech recognition embedding and disentanglement losses, in: ICASSP 2022-2022 IEEE International Conference on Acoustics, Speech and Signal Processing (ICASSP), IEEE, 2022, pp. 391--395.

\bibitem{du2020self}
Z.~Du, M.~Lei, J.~Han, S.~Zhang, Self-supervised adversarial multi-task learning for vocoder-based monaural speech enhancement., in: Interspeech, 2020, pp. 3271--3275.

\bibitem{ma2018modeling}
J.~Ma, Z.~Zhao, X.~Yi, J.~Chen, L.~Hong, E.~H. Chi, Modeling task relationships in multi-task learning with multi-gate mixture-of-experts, in: Proceedings of the 24th ACM SIGKDD international conference on knowledge discovery \& data mining, 2018, pp. 1930--1939.

\bibitem{han2022dynamic}
Y.~Han, G.~Huang, S.~Song, L.~Yang, H.~Wang, Y.~Wang, Dynamic neural networks: A survey, IEEE Transactions on Pattern Analysis and Machine Intelligence 44~(11) (2022) 7436--7456.
\newblock \href {https://doi.org/10.1109/TPAMI.2021.3117837} {\path{doi:10.1109/TPAMI.2021.3117837}}.

\bibitem{JesperK2023PSD}
J.~K. Nielsen, M.~G. Christensen, J.~B. Boldt, An analysis of traditional noise power spectral density estimators based on the gaussian stochastic volatility model, IEEE/ACM Transactions on Audio, Speech, and Language Processing 31 (2023) 2299--2313.
\newblock \href {https://doi.org/10.1109/TASLP.2023.3282107} {\path{doi:10.1109/TASLP.2023.3282107}}.

\bibitem{wang2020glance}
Y.~Wang, K.~Lv, R.~Huang, S.~Song, L.~Yang, G.~Huang, Glance and focus: A dynamic approach to reducing spatial redundancy in image classification, in: Proceedings of the 34th International Conference on Neural Information Processing Systems, NIPS'20, Curran Associates Inc., Red Hook, NY, USA, 2020.

\bibitem{kirillov2020pointrend}
A.~Kirillov, Y.~Wu, K.~He, R.~Girshick, Pointrend: Image segmentation as rendering, in: Proceedings of the IEEE/CVF conference on computer vision and pattern recognition, 2020, pp. 9799--9808.

\bibitem{elbayad2019depth}
M.~Elbayad, J.~Gu, E.~Grave, M.~Auli, Depth-adaptive transformer, arXiv preprint arXiv:1910.10073 (2019).

\bibitem{hansen2019neural}
C.~Hansen, C.~Hansen, S.~Alstrup, J.~G. Simonsen, C.~Lioma, Neural speed reading with structural-jump-lstm, arXiv preprint arXiv:1904.00761 (2019).

\bibitem{tavarone2018conditional}
R.~Tavarone, L.~Badino, Conditional-computation-based recurrent neural networks for computationally efficient acoustic modelling., in: Interspeech, 2018, pp. 1274--1278.

\bibitem{wang2019tafe}
X.~Wang, F.~Yu, R.~Wang, T.~Darrell, J.~E. Gonzalez, Tafe-net: Task-aware feature embeddings for low shot learning, in: Proceedings of the IEEE/CVF conference on computer vision and pattern recognition, 2019, pp. 1831--1840.

\bibitem{kang_incorporating_2020}
D.~Kang, D.~Dhar, A.~B. Chan, \href{https://doi.org/10.1007/s11263-020-01345-8}{Incorporating {Side} {Information} by {Adaptive} {Convolution}}, International Journal of Computer Vision 128~(12) (2020) 2897--2918.
\newblock \href {https://doi.org/10.1007/s11263-020-01345-8} {\path{doi:10.1007/s11263-020-01345-8}}.
\newline\urlprefix\url{https://doi.org/10.1007/s11263-020-01345-8}

\bibitem{iwamoto2022bad}
K.~Iwamoto, T.~Ochiai, et~al., How bad are artifacts?: Analyzing the impact of speech enhancement errors on asr, arXiv preprint arXiv:2201.06685 (2022).

\bibitem{le2019sdr}
J.~Le~Roux, S.~Wisdom, et~al., Sdr--half-baked or well done?, in: Proc.\ IEEE Int.\ Conf.\ Acoust., Speech, Signal Process., IEEE, 2019, pp. 626--630.

\bibitem{vincent2006performance}
E.~Vincent, R.~Gribonval, C.~F{\'e}votte, Performance measurement in blind audio source separation, {IEEE} Audio, Speech, Language Process. 14~(4) (2006) 1462--1469.

\bibitem{wang2023harmonic}
T.~Wang, W.~Zhu, et~al., Harmonic attention for monaural speech enhancement, {IEEE/ACM} Trans. Audio, Speech, Language Process. (2023).

\bibitem{burchi2021efficient}
M.~Burchi, V.~Vielzeuf, Efficient conformer: Progressive downsampling and grouped attention for automatic speech recognition, in: 2021 IEEE Automatic Speech Recognition and Understanding Workshop (ASRU), IEEE, 2021, pp. 8--15.

\bibitem{Chung2018VoxCeleb2DS}
J.~S. Chung, A.~Nagrani, A.~Zisserman, \href{https://api.semanticscholar.org/CorpusID:49211906}{Voxceleb2: Deep speaker recognition}, in: Interspeech, 2018.
\newline\urlprefix\url{https://api.semanticscholar.org/CorpusID:49211906}

\bibitem{desplanques20_interspeech}
B.~Desplanques, J.~Thienpondt, K.~Demuynck, {ECAPA-TDNN: Emphasized Channel Attention, Propagation and Aggregation in TDNN Based Speaker Verification}, in: Proc. Interspeech 2020, 2020, pp. 3830--3834.
\newblock \href {https://doi.org/10.21437/Interspeech.2020-2650} {\path{doi:10.21437/Interspeech.2020-2650}}.

\bibitem{hsu2021hubert}
W.~Hsu, B.~Bolte, et~al., Hubert: Self-supervised speech representation learning by masked prediction of hidden units, {IEEE/ACM} Trans. Audio, Speech, Language Process. 29 (2021) 3451--3460.

\bibitem{dubey2022icassp}
H.~Dubey, V.~Gopal, et~al., Icassp 2022 deep noise suppression challenge, in: Proc.\ IEEE Int.\ Conf.\ Acoust., Speech, Signal Process., IEEE, 2022, pp. 9271--9275.

\bibitem{kingma2014adam}
D.~P. Kingma, J.~Ba, Adam: A method for stochastic optimization, arXiv preprint arXiv:1412.6980 (2014).

\bibitem{recommendation2001perceptual}
I.-T. Recommendation, Perceptual evaluation of speech quality (pesq): An objective method for end-to-end speech quality assessment of narrow-band telephone networks and speech codecs, Rec. ITU-T P. 862 (2001).

\bibitem{taal2011algorithm}
C.~H. Taal, R.~C. Hendriks, et~al., An algorithm for intelligibility prediction of time--frequency weighted noisy speech, {IEEE} Audio, Speech, Language Process. 19~(7) (2011) 2125--2136.

\bibitem{reddy2021dnsmos}
C.~K. Reddy, V.~Gopal, R.~Cutler, Dnsmos: A non-intrusive perceptual objective speech quality metric to evaluate noise suppressors, in: Proc.\ IEEE Int.\ Conf.\ Acoust., Speech, Signal Process., IEEE, 2021, pp. 6493--6497.

\end{thebibliography}





\end{document}